\documentclass[final,1p,times,english]{elsarticle}

\usepackage[T1]{fontenc}
\usepackage[latin9]{inputenc}
\usepackage{fancybox}
\usepackage{calc}
\usepackage{mathrsfs}
\usepackage{amsmath}
\usepackage{amssymb}
\usepackage{graphicx}
\usepackage{setspace}
\usepackage{subfigure}
\usepackage[bold]{hhtensor}
\usepackage{hyperref}
\hypersetup{
  colorlinks,
  citecolor=blue,
  linkcolor=blue,
  urlcolor=blue}

\newcommand\blfootnote[1]{%
  \begingroup
  \renewcommand\thefootnote{}\footnote{#1}%
  \addtocounter{footnote}{-1}%
  \endgroup
}

\makeatletter

\DeclareRobustCommand{\greektext}{%
  \fontencoding{LGR}\selectfont\def\encodingdefault{LGR}}
\DeclareRobustCommand{\textgreek}[1]{\leavevmode{\greektext #1}}
\DeclareFontEncoding{LGR}{}{}
\DeclareTextSymbol{\~}{LGR}{126}

\makeatother

\usepackage{babel}
\begin{document}
\begin{frontmatter}

\title{Computationally-efficient stochastic cluster dynamics method for modeling damage accumulation
in irradiated materials}

\author[a1,a2]{Tuan L. Hoang}
\author[a2,a3]{Jaime Marian\blfootnote{Corresponding author: Jaime Marian \\
\-\hspace{0.5cm}email: \href{mailto:jmarian@ucla.edu}{jmarian@ucla.edu}}}
\author[a2]{Vasily V. Bulatov}
\author[a1]{Peter Hosemann}

\address[a1]{Department of Nuclear Engineering, University of California,
Berkeley, CA 94720, USA}
\address[a2]{Physical and Life Sciences Directorate, Lawrence Livermore National Laboratory, CA 94550, USA}
\address[a3]{Department of Materials Science and Engineering, University
of California, Los Angeles, CA 94720, USA}

\begin{abstract}
\begin{doublespace}
An improved version of a recently developed stochastic cluster dynamics
(SCD) method {[}Marian, J. and Bulatov, V. V., {\it J. Nucl. Mater.} \textbf{415} (2014)
84-95{]} is introduced as an alternative to rate theory (RT) methods
for solving coupled ordinary differential equation (ODE) systems for
irradiation damage simulations. SCD circumvents by design the curse
of dimensionality of the variable space that renders traditional ODE-based
RT approaches inefficient when handling complex defect population
comprised of multiple (more than two) defect species. Several improvements
introduced here enable efficient and accurate simulations of irradiated
materials up to realistic (high) damage doses characteristic of next-generation
nuclear systems. The first improvement is a procedure for efficiently
updating the defect reaction-network and event selection in the context
of a dynamically expanding reaction-network. Next is a novel implementation
of the $\tau$-leaping method that speeds up SCD simulations by advancing
the state of the reaction network in large time increments when appropriate.
Lastly, a volume rescaling procedure is introduced to control the
computational complexity of the expanding reaction-network through
occasional reductions of the defect population while maintaining accurate
statistics. The enhanced SCD method is then applied to model defect
cluster accumulation in iron thin films subjected to triple ion-beam
($\text{Fe}^{3+}$, $\text{He}^{+}$ and $ $$\text{H\ensuremath{{}^{+}}}$$ $)
irradiations, for which standard RT or spatially-resolved kinetic
Monte Carlo simulations are prohibitively expensive. 
\end{doublespace}

\end{abstract}

\end{frontmatter}

\begin{doublespace}

\section{Introduction}
\end{doublespace}

\begin{doublespace}
The production and accumulation of defects in materials subjected
to irradiation is a multiscale problem spanning multiple orders of
magnitude in time and space. For the last several decades, the rate
theory (RT) method for solving coupled ordinary differential equation
(ODE) systems has been the workhorse for irradiation damage simulations \cite{bull,ghoniem,mansur1996},
mostly owing to its much greater computational efficiency compared
to more detailed methods such as molecular dynamics (MD) or kinetic
Monte Carlo (kMC). RT involves solving a set of coupled ODEs such
as:

\begin{equation}
\dfrac{dC_{i}}{dt}=\overset{.}{\mathcal{F}_{i}}-\overset{.}{\mathcal{L}_{i}},\qquad(i=1,...,N)\label{eq:1}
\end{equation}
where each equation describes the time evolution of the average concentration
of a particular type (species) of defect cluster denoted by index
$i$. The terms on the right hand side are the loss rate $\overset{.}{\mathcal{L}_{i}}$
of species $i$ due to various kinetic processes, and the production
rate $\overset{.}{\mathcal{F}_{i}}$ of species $i$ due to irradiation
and reactions involving defect cluster species other than $i$. RT
models achieve a high level of simulation efficiency at the cost of
drastic simplifications in the underlying physical model, chief of
which is the mean-field approximation that neglects spatial correlations
and finite volume fluctuations. Another significant reduction in computational
complexity is gained by limiting the number of species considered.
In practice, the number of admissible defect species (and ODEs in
the system) is truncated to achieve a satisfactory balance between
accuracy and available computational resources. Large defect clusters
not explicitly included in the set are accounted for only approximately
(if at all) using a truncation model for the tail of the defect size
distribution \cite{koiwa1974,sands2013}
\footnote{Existing truncation schemes are ad hoc and unlikely to correctly capture
the statistic of extreme values in the defect size distribution believed
to be important for understanding material degradation under irradiation
 }. Once defined, the number of ODEs in the set must remain the same
through the simulation. To allow simulations to realistically high
irradiation doses, this number may need to be as high as $10^{6}$
even in the simplest materials, e.g.~pure iron. Furthermore, the number
of distinct ODEs that need to be included in the set grows exponentially
with increasing number of complex defect cluster types, e.g. simulations
of $\text{V}_m\text{He}_n$ complexes of
$m$ vacancies and $n$ helium atoms requires $(m\times n)$
equations to be included. This is yet another case of combinatorial
explosion where the number of equations to be solved is far too large
for practical numerical simulations. Consequently, current RT models
have been limited to defect populations having no more than two and,
in most cases, only one size dimension. This need to allocate an ODE
for every possible defect cluster type even before the simulation
starts is a serious limitation of the ODE-based RT method. 

To overcome these limitations, Marian and Bulatov recently developed
the stochastic cluster dynamics (SCD) method to model defect evolution
in irradiated materials \cite{marian2011}. The SCD method is based on the stochastic
simulation algorithm (SSA) proposed originally by Gillespie for simulations
of chemical kinetics in well-stirred systems \cite{gillespie1977,asher}. Whereas RT is
formulated in terms of average species concentrations that can take
arbitrary fractional values, SSA considers integer-valued species
populations in a finite volume and interprets the ODEs defining the
RT model as a set of stochastic master equations. The so-defined species
population is then evolved stochastically, one reaction at a time,
following a standard kMC algorithm. The SSA method has been widely
used in the chemical engineering and biochemistry communities \cite{gillespie1992,cao2004,cai2007,ahn2008,cain}
but is still relatively unknown to computational materials scientists.
SCD achieves additional efficiency through the use of dynamic data
handling mechanisms where only defect clusters with non-zero populations
are kept track of throughout the simulation time. This is a major
advantage over RT in which every admissible defect cluster must be
allocated a variable and an equation that persist through all stages
of ODE integration. Importantly, the computational complexity of a
SCD simulation is controlled by the value of the simulation volume
and does not depend on the complexity (number of size dimensions)
of admissible defect cluster types. Thus, SCD does not suffer from
combinatorial explosion and can handle cluster populations with arbitrary
number of size attributes. Several proof-of-principle studies have
been carried out to demonstrate the applicability of the SCD method
to simulations of irradiated materials \cite{marian2011}.

Although SCD sidesteps combinatorial explosion, the method relies
on a kMC algorithm to sample stochastic evolution trajectories from
the master equation. Thus, SCD simulations face the usual computational
challenges characteristic of kMC simulation methods, such as stiffness
caused by a wide spectrum of event rates. Further applications of
SCD to technologically relevant materials and irradiation conditions
require improvements to make the method more robust and computationally
efficient. In this paper, we present several enhancements to SCD,
specifically (i) a dynamic reaction-network expansion mechanism to
efficiently update the reaction channels and the total reaction rate,
(ii) an implementation of the $\tau$-leaping algorithm to accelerate
SCD simulations by allowing several reaction events to be leaped over
in one single time-step $\tau$, and (iii) a volume scaling method
in which the reaction volume is reduced adaptively in order to control
the computational cost while preserving statistically significant
defect populations. The $\tau$-leaping method \cite{tauleap} was originally
developed and used in SSA simulations with fixed variable spaces \cite{cain}.
In SCD, where the size of the reaction network varies with time, an
efficient algorithm for updating noncritical reactions and noncritical
species and for computing the leap time is needed to reduce the overhead
associated with $\tau$-leaping.  We apply the enhanced SCD method
to simulations of defect populations in pure iron subjected to triple
ion-beam irradiation. The predicted damage accumulation kinetics are
verified by comparing them to the original SCD algorithm predictions.
The same comparisons are used to quantify gains in computational performance
over the original SCD simulations.

The paper is organized as follows. In Section \ref{back}, we overview the theory
behind the SSA and the $\tau$-leaping methods. In Section \ref{scd}, we briefly
overview our original SCD algorithm, our material model for iron and
the types of reaction events considered in our radiation damage simulations.
Improvements to the SCD method are described in Section \ref{improved} together
with their algorithmic details. In Section \ref{numero}, we present the numerical
verification of the new improved SCD algorithm and compare its computational
performance to the original algorithm. Finally, Section \ref{conc} summarizes
our findings.
\end{doublespace}

\begin{doublespace}

\section{Background}\label{back}
\end{doublespace}

\begin{doublespace}

\subsection{The Stochastic Simulation Algorithm (SSA)}
\end{doublespace}

\begin{doublespace}
For clarity, we briefly summarize the SSA method developed by Gillespie
for simulations of chemical reactions in well-stirred systems. The
reader is referred to the original paper \cite{gillespie1977} for more details
of the method and the theory behind it. Consider a population containing
$N$ defect-clusters ${S_{1},S_{2},..,S_{N}}$ that can participate
in $M$ reaction channels $\{\mathbb{R}_{1},\mathbb{R}_{2},...,\mathbb{R}_{M}\}$.
Let $\overrightarrow{X}(t)$ be the dynamic state vector of the system
at an arbitrary time $t$, $\overrightarrow{X}(t)=\{X_{1}(t),X_{2}(t),\text{\ldots},X_{N}(t)\}$,
where $X_{i}(t)$ is the number of defect clusters of type $S_{i}$
at time $t$. Each reaction channel is characterized by its reaction
rate $R_{j}$ and by its state change vector $\mathbf{\overrightarrow{\nu}}_{j}$
= ($\mathbf{\nu}_{1j}$, $\mathbf{\nu}_{2j}$, \ldots{}, $\mathbf{\nu}_{Mj}$).
The probability that a reaction of type $j$ will take place within
the next infinitesimal time interval $[t,t+dt)$ is given by the product
$R_{j}dt$ whereas $\nu_{ij}$ specifies the change in the population
of species $S_{i}$ after a single reaction event along channel $\mathbb{R}_{j}$.
The evolution of such reaction network obeys the following chemical
master equation (CME)

\begin{equation}
\dfrac{\partial P(\overrightarrow{x},t|\overrightarrow{x_{0}},t_{0})}{\partial t}=\sum_{j=1}^{M}[R_{j}(\overrightarrow{x}-\overrightarrow{\mathbf{\nu}_{j}})P(\overrightarrow{x}-\overrightarrow{\mathbf{\nu}_{j}},t|\overrightarrow{x_{0}},t_{0})-R_{j}(\overrightarrow{x})P(\overrightarrow{x},t|\overrightarrow{x_{0}},t_{0})]\label{eq:1-1}
\end{equation}
where $P(\overrightarrow{x},t|\overrightarrow{x_{0}},t_{0})$ is the
conditional probability that $\overrightarrow{X}(t)=\overrightarrow{x}$
at time $t$ if $\overrightarrow{X}(t_{0})=\overrightarrow{x_{0}}$
at time $t_{0}$. The above CME defines a stochastic process referred
to as a continuous time Markov chain. Rather than attempting to solve
this CME equation directly, individual stochastic time trajectories
of the state vector $\overrightarrow{X}(t)$ can be obtained using
an appropriate kinetic Monte Carlo algorithm. In particular, in the
following algorithm two random numbers $r_{1}$ and $r_{2}$ uniformly
distributed in $[0,1)$ are generated. The time to the next reaction
event is then given by

\begin{equation}
\Delta t=-\dfrac{1}{\sum_{j}R_{j}}log\left(\dfrac{1}{r_{1}}\right),\label{eq:dt}
\end{equation}
and the index of the same reaction event, $R_{k}$, is taken to be
he smallest integer $k$ that satisfies the following condition 
\begin{equation}
\sum_{i=1}^{k}R_{i}>r_{2}\sum_{j=1}^{M}R_{j}=r_{2}R_{tot}.\label{eq:random variate}
\end{equation}
where $R_{tot}=\sum^M_jR_j$. Once the
next reaction event and its time increment are selected, the simulation
time and the state vector are updated accordingly, $t=t_{0}+\Delta t$
and $\overrightarrow{X}(t+\Delta t)=\overrightarrow{X}(t_{0})+\overrightarrow{\nu_{j}}$.
The simulation proceeds to the next reaction event until the desired
simulation time is reached. The method just described is referred
to as \emph{direct} SSA method. The direct SSA method rigorously
generates stochastic trajectories sampled for the exact (even if often
unknown) solutions of the CME. Several algorithmic enhancements have
been proposed to improve efficiency of the direct SSA method,
including the \emph{first reaction} method \cite{gillespie1977}, the \emph{modified}
direct method \cite{cao2004}, the \emph{optimized} direct method \cite{gibson2000},
the \emph{sorting} direct method \cite{mac2006}, or the \emph{logarithmic}
direct method \cite{petzold2006}, to name a few. Any such improvements notwithstanding,
simulating every reaction event one at a time is often impractical
for large reaction networks of practical interest. To address this
problem, Gillespie proposed the $\tau$-leaping method that allows
many reactions channels to fire in a single timestep at the expense
of some minor accuracy loss. Because conditions that justify the using
of $\tau$-leaping are often met in radiation damage simulations,
in the following we briefly describe $\tau$-leaping as a way to accelerate
stochastic simulations.
\end{doublespace}

\begin{doublespace}

\subsection{The $\tau$-leaping method}
\end{doublespace}

\begin{doublespace}
The $\tau$-leaping method is based on the leap condition which assumes
that a reaction channel may be fired multiple times within a small time interval $[t, t + \tau)$ if the reaction rate does not suffer significant changes over that interval.
Then, given the state vector of the system $\overrightarrow{X}(t)=\overrightarrow{x}$,
the number of times that each reaction channel $\mathbb{R}_{j}$ can 
fire is approximated by the Poisson distribution $\mathcal{P}\left(R_{j}\tau\right)$.
The simulation proceeds as follows: (i) at each time-step we find
a value of $\tau$ that satisfies the leap condition mentioned above;
(ii) for each $\nu_{j}$, a Poisson random number with mean $R_{j}\tau$,
i.e. $\mathcal{P}\left(R_{j}\tau\right)$ is generated; (iii) the
system is updated as $\overrightarrow{X}(t+\tau)\leftarrow\overrightarrow{X}(t)+\sum_{j}^{M}\mathcal{P}\left(R_{j}\tau\right)\nu_{j}$,
and the simulation time advances to the new time $t\leftarrow t+\tau$.
As a result, the simulation can be accelerated at a greater speed
since the it can leap through multiple reactions in one single
step instead of firing the reactions one by one.
\end{doublespace}

\begin{doublespace}

\section{The Stochastic Cluster Dynamics algorithm}\label{scd}
\end{doublespace}

\begin{doublespace}

\subsection{Model representation}
\end{doublespace}

\begin{doublespace}
At any point in time the state of the model is characterized by the
set of all existing clusters $\overrightarrow{S}_{all}=\left\{ S_{i}\right\} $.
Dynamic updates of state vectors are efficiently handled using hash
tables with dynamic resizing. More details on the hash functions and
associated operations are given in the next section. Each cluster
$S_{i}$ contains several associated attributes such as the number
of each component species contained in the cluster, the cluster species
population count, its diffusion coefficient, the binding energies
among the component subspecies and the cluster, and other relevant
parameters. Mobile species with a nonzero diffusivity are regarded
as a subset $\overrightarrow{S}_{m}$ (here $m$ stands for mobile)
of $\overrightarrow{S}_{all}$. Defect
cluster species associated with a recently executed event are stored
in a dynamic array whose purpose will be described in the following
section. Such species can be reactants or products of a recently executed
reaction event or a collection of defects and clusters that have just
been introduced into the volume as a result of a defect insertion
event (due to irradiation). The evolving reaction network $\overrightarrow{\mathcal{\mathscr{\mathbb{\boldsymbol{\mathtt{\mathfrak{\mathscr{\mathbb{R}}}}}}}}}=\left\{ \mathbf{\mathscr{\mathbb{R}}}_{i}\right\} $
specifies all reaction channels available for the current defect population
$\overrightarrow{S}_{all}$. Each binary reaction channel $\mathbf{\mathscr{\mathbb{R}}}(S_{1},S_{2})$
represents a reaction between species of type $S_{1}$ and type $S_{2}$
with an associated reaction rate $R(S_{1},S_{2})$ (clusters $S_{1}$
and $S_{2}$ can be identical when the reaction involves two like
species). To implement the $\tau$-leaping method, two more data sets
will be defined. The first set $\overrightarrow{\mathcal{\mathtt{\mathbb{J}}}}=\left\{ \mathbf{\mathbb{J}}_{i}\right\} $
contains all noncritical reaction channels whose associated reactants
have populations larger than a certain user-predefined value $n_{cr}$.
Another set $\overrightarrow{P}=\left\{ P_{i}\right\} $ contains
all defect cluster species associated with the noncritical reactions.
Each $P_{i}$ contains a parameter specifying the highest order of
possible reactions species $i$ can participate in, as explained in
more details later these reaction order parameters are utilized in
computing the leap time $\tau$.
\end{doublespace}

\begin{doublespace}

\subsection{Types of events}
\end{doublespace}

\begin{doublespace}
Hereafter, $\text{V}_s$ and $\text{I}_s$ denote
a vacancy cluster or a self interstitial atom (SIA) cluster of size
$s$. In our model, we only consider clusters with a maximum of three
component species, specifically the clusters only contain He and H
atoms together with either vacancies or interstitials of the host
material. If desired, the model can be modified to admit defect clusters
of arbitrarily complex compositions. The following reactions are currently
admitted in our SCD model of iron:
\end{doublespace}

\noindent \begin{flushleft}
\doublebox{\begin{minipage}[t]{1\columnwidth}%
\begin{description}
\item [{\textmd{$0^{\text{th}}$-order}}] reactions\end{description}
\begin{itemize}
\item \begin{flushleft}
Defect insertion, e.g. generation of certain types of defects resulting
from collisions of incoming energetic particles with the host matrix
atoms. 
\par\end{flushleft}
\end{itemize}
$1^{\text{st}}$-order reactions
\begin{itemize}
\item \begin{flushleft}
Defect absorption at sinks: mobile clusters can migrate towards sinks
and become absorbed there. Sinks can be free surfaces, dislocation
networks or grain boundaries.
\par\end{flushleft}
\item \begin{flushleft}
Emission of a monomer from a defect cluster: a cluster can emit a
monomer of one of its constituent species, reducing its species
count appropriately. A complex cluster cluster $\text{V}_i\text{He}_j\text{H}_k$
can emit a vacancy, or a He monomer or a H monomer. Following emission,
the initial cluster's population is reduced by one and two new defect
species are created or, if one or both species already exist, their
counts are increased by one. For example, emission of one vacancy
V (or one He monomer) produces a smaller defect cluster $\text{V}_{(i-1)}\text{He}_j\text{H}_k$
(or $\text{V}_i\text{He}_{(j-1)}\text{H}_k$
in case the monomer is a He atom).
\par\end{flushleft}
\end{itemize}
$2^{\text{nd}}$-order reactions
\begin{itemize}
\item \begin{flushleft}
Defect annihilation: collisions of two clusters containing vacancies
and self interstitial atoms result in their complete or partial recombination.
For example, collision of a complex vacancy cluster $\text{V}_i\text{He}_j\text{H}_k$
with a SIA cluster $\text{I}_{i^{\prime}}$ produces $\text{V}_{(i-i^{\prime})}\text{He}_j\text{H}_k$
(if $i>i^{\prime}$) or $\text{I}_{\left(i^{\prime}-i\right)}\text{He}_j\text{H}_k$
(if $i^{\prime}>i$ ) or releases $ $$j$ monomers
of $\text{He}$ and $k$ monomers of $\text{H}$ monomers (if
$i=i^{\prime}$). He and H monomers are assumed not
to bind unless vacancies and interstitials are also present.
\par\end{flushleft}
\item \begin{flushleft}
Defect aggregation: clusters containing like defects can combine to
form larger clusters upon interaction. For example, a $\text{V}_i\text{He}_j\text{H}_k$
cluster can collide with a $\text{V}_{i^{\prime}}\text{He}_{j^{\prime}}\text{H}_{k^{\prime}}$
cluster producing a larger $\text{V}_{(i+i^{\prime})}\text{He}_{(j+j^{\prime})}\text{H}_{(k+k^{\prime})}$
cluster. 
\par\end{flushleft}\end{itemize}
\end{minipage}}
\par\end{flushleft}

\begin{doublespace}

\subsection{Summary of the original SCD algorithm}
\end{doublespace}

\begin{doublespace}
The main motivation for the development of SCD was to avoid combinatorial
explosion in the number of equations encountered in traditional ODE-based
RT simulations. In SCD, the simulation volume is finite and defect
cluster species have integer-valued populations. In a typical initial
state, relatively few (if any) defect species exist, so, rather than
allocating memory for all possible defect-clusters before the start
of the simulation, cluster species are added or removed from the hash
table dynamically, as needed. Therefore, only defect clusters that
have nonzero populations are kept track of. The hash table is implemented
as an associative array in which a hash function is used to map the
identifying values --known as hash \emph{keys}-- to their associated
values. The hash function maps the keys onto the index array elements
(known as buckets) where the associated values are stored. Operations on
a hash table such as adding, removing or locating buckets take constant
time on average and do not depend of the size of the hash itself unlike
operations on indexed arrays. In simulations of irradiated materials, the number of pre-existing defect clusters is usually
small but increases rapidly after high energy particles begin to create defects.
 Furthermore, defect populations and their associated reaction channels
change with each subsequent reaction event. It is our experience
that in such conditions hashing is more efficient than using array structures
for handling large and evolving data sets since defect clusters can
be located and updated quickly. The original SCD algorithm consists
of the following steps: 
\end{doublespace}

\begin{singlespace}
\noindent \begin{flushleft}
\doublebox{\begin{minipage}[t]{1\columnwidth}%
\begin{enumerate}
\item \begin{flushleft}
Construct two hash tables: one, $\overrightarrow{S}_{all}$, to store
all the existing defect clusters and another one, $\overrightarrow{S}_{m}$,
to store only the mobile defects in $\overrightarrow{S}_{all}$, . 
\par\end{flushleft}
\item \begin{flushleft}
Construct a reaction table $\overrightarrow{\mathscr{\mathbb{R}}}$
containing the reaction channels involving all existing defect clusters,
and store $\overrightarrow{\mathscr{\mathbb{R}}}$ in an array. 
\par\end{flushleft}
\item \begin{flushleft}
Calculate the total reaction rate by summing the rates of all currently
existing reaction channels in the reaction table.
\par\end{flushleft}
\item \begin{flushleft}
Randomly select the time increment to the next reaction as well as
the type of the reaction event using eqs. \ref{eq:dt} and \ref{eq:random variate}.
\par\end{flushleft}
\item \begin{flushleft}
Execute the selected reaction event, update the hash tables accordingly
and delete the reaction table $\overrightarrow{\mathscr{\mathbb{R}}}$.
\par\end{flushleft}
\item \begin{flushleft}
Return to step 2 and proceed until the total simulation time is reached.
\par\end{flushleft}\end{enumerate}
\end{minipage}}
\par\end{flushleft}
\end{singlespace}

\begin{doublespace}
Using an array to store the reaction channels $\overrightarrow{\mathscr{\mathbb{R}}}$
proves to be inefficient due to the highly dynamic nature of stochastic
evolution. Furthermore it is wasteful to build the reaction table
anew after every reaction event since only a portion of the reaction
channels is changed due to the executed event while most others are
left intact. These two inefficiencies are addressed in the improved
version of SCD presented in the following section.
\end{doublespace}

\begin{doublespace}

\section{An improved Stochastic Cluster Dynamics algorithm}\label{improved}
\end{doublespace}

\begin{doublespace}
Except for massive defect insertion events representing collision
cascades, only a small number of defect clusters in the simulation
volume are affected by a single reaction event. Therefore, only the
reaction channels involving affected defect species need to be updated,
while the rest of the reaction network remains untouched. In this
enhanced version of SCD, we use hashing to maintain existing species
and reaction channels and to expand the reaction network when new
species are introduced by the reaction events. Such updates are typically
more efficient than the reconstruction of the entire reaction table in between insertion events. Depending on the specific reaction
model implemented, some defect species become quite numerous and their
associated reaction channels can fire much more frequently than others.
For example in our model for iron, SIAs and vacancies are observed
to migrate in large numbers to defects sinks soon after irradiation
commences, whereas defect insertion and defect association events are
relatively infrequent. To expedite SCD simulations under such conditions,
we implement a version of $\tau$-leaping method in which several
repetitive reaction events are executed at once. Lastly, we introduce
and justify a volume rescaling procedure to reduce the computational
complexity of SCD simulations at later stages of damage accumulation.
This is when the density of defect clusters becomes high, and the diffusion
length of mobile defects becomes small compared to the linear dimension
of the simulation volume.
\end{doublespace}

\begin{doublespace}

\subsection{Dynamic reaction network updates and expansion}\label{dynnetwork}
\end{doublespace}

\begin{doublespace}
As follows from eqs.~\ref{eq:dt} and \ref{eq:random variate}, both the time increment to the next
reaction event and the type of reaction are selected based on the
total event rate summed over all existing reaction channels. In the
original version of SCD, the net event rate was recomputed after each
reaction event throughout the simulation. However, in a production
scale SCD simulation the number of distinct reaction rates grows rapidly
to thousands and even millions and yet only a small sub-set of reaction
channels is directly affected by each reaction event. Enabling incremental
updates requires that reaction channels affected (modified or eliminated)
by the last event be located and updated in the computer's memory
efficiently during the course of the simulation. We rely on hashing
to quickly add, remove, locate and update reaction channels in real
time.

In the improved version of SCD reported here, in addition to the two
hash tables $\overrightarrow{S}_{all}$ and $\overrightarrow{S}_{m}$
used to store and reference the total and the mobile cluster populations,
all existing reaction channels are stored in a reaction hash $\overrightarrow{\mathscr{\mathbb{R}}}$.
The reaction table expands or contracts as needed to accommodate new
reactions associated with the creation (or extinction) of new defect
species. The process for updating the affected hash tables goes as
follows: 
\end{doublespace}
\begin{itemize}
\begin{doublespace}
\item A new hash key is created for all possible species resulting from
these reactions. For each 2$^{{\rm nd}}$-order reaction, this key
is generated from the keys of its constituent reactants stored in
$\overrightarrow{S}_{all}$ while for 0$^{{\rm th}}$ and 1$^{{\rm st}}$-order
reactions, dummy keys --two and one, respectively-- are used as appropriate. 
\item Each cluster in the $\overrightarrow{S}_{all}$ hash table is assigned
a parameter $f_{1}$ indexing its count change due to the recently
executed event; another parameter $f_{2}$ indicates whether the defect
already existed in the simulation volume in the previous time-step.
These parameters let SCD know whether it should look up and update
the existing reaction $\mathbb{R}(S_{1},S_{2})$ or add it as a newly
created one into the reaction hash table $\overrightarrow{\mathscr{\mathbb{R}}}$. 
\item As a cost-savings measure, defect clusters that have participated
in a recent reaction event are stored in a dynamic array so that product
species can be updated efficiently. As the number of these clusters
is not very large, a dynamic array is simpler than a hash table in
this case. \end{doublespace}

\end{itemize}
\begin{doublespace}

\subsection{Reaction rate updating}\label{update}
\end{doublespace}

\begin{doublespace}
The first step of the reaction update process is to visit each reaction
in $\overrightarrow{\mathscr{\mathbb{R}}}$ and remove those whose
component reactants no longer exist due to the previous event(s).
Subsequently, we visit each element $S_{i}$ in $\overrightarrow{S}_{all}$
and update all the reaction channels that this cluster associates
with. As a result, some existing reactions in $S_{i}$ will be modified
and new reactions will be added into the reaction hash table $\overrightarrow{\mathscr{\mathbb{R}}}$.

Based on the values of $f_{1}$ and $f_{2}$ mentioned previously,
it can be established whether a cluster was a reactant or product
of the last reaction event. If $S_{i}$ is a new defect cluster, all
the reactions associated with it will be added directly into the reaction
hash table because it is not necessary to check for their existence
in it. On the other hand, if the cluster $S_{i}$ only increases or
decreases in number, all of its associated reactions will be first
located in the reaction hash and updated accordingly based on the
value of $f_{1}$. If $f_{1}$ is the change in population of cluster
$S_{i}$ and $R$ is the rate of a reaction channel involving $S_{i}$,
then the total reaction rate $R_{tot}$ can be updated as follows:

$\text{1}^{\text{st}}$-order reaction: 
\begin{equation}
R(S_{i})\leftarrow R_{0}(S_{i})\left[1+\frac{f_{1}(S_{i})}{\mbox{\ensuremath{X_{0}(S_{i})}}}\right],\qquad R_{tot}\leftarrow R_{tot}+R_{0}(S_{i})\dfrac{f_{1}(S_{i})}{X_{0}(S_{i})}\label{eq:3}
\end{equation}

For $2^{\text{nd}}$-order reactions between a cluster $S_{i}$ and
another cluster $S_{j}$ (assuming $\mathbb{R}(S_{i},S_{j})$ already
exists): 
\begin{equation}
\begin{cases}
R(S_{i},S_{j})\leftarrow R_{0}(S_{i},S_{j})\left[1+\frac{f_{1}(S_{i})}{\mbox{\ensuremath{\mbox{\ensuremath{X_{0}(S_{i})}}}}}\right]\left[1+\frac{f_{1}(S_{j})}{\mbox{\ensuremath{X_{0}(S_{j})}}}\right] & (S_{i}\neq S_{j})\\
R(S_{i},S_{j})\leftarrow R_{0}(S_{i},S_{j})\left[1+\frac{f_{1}(S_{i})}{\mbox{\ensuremath{\mbox{\ensuremath{X_{0}(S_{i})}}}}}\right]\left[1+\frac{f_{1}(S_{i})}{\mbox{\ensuremath{\mbox{\ensuremath{X_{0}(S_{i})}}-1}}}\right] & (S_{i}\equiv S_{j})
\end{cases}\label{eq:4}
\end{equation}
\begin{equation}
\begin{cases}
R_{tot}\leftarrow R_{tot}+R_{0}(S_{i},S_{j})\mbox{\ensuremath{\left[\mbox{\ensuremath{\frac{f_{1}(S_{i})}{\mbox{\ensuremath{\mbox{\ensuremath{X_{0}(S_{i})}}}}}}+\ensuremath{\frac{f_{1}(S_{j})}{\mbox{\ensuremath{\mbox{\ensuremath{X_{0}(S_{j})}}}}}}+\ensuremath{\frac{f_{1}(S_{i})f_{1}(S_{j})}{\mbox{\ensuremath{\mbox{\ensuremath{X_{0}(S_{i})X_{0}(S_{j})}}}}}}}\right]}} & (S_{i}\neq S_{j})\\
R_{tot}\leftarrow R_{tot}+R_{0}(S_{i},S_{j})\mbox{\ensuremath{\left[\mbox{\ensuremath{\frac{f_{1}(S_{i})}{\mbox{\ensuremath{\mbox{\ensuremath{X_{0}(S_{i})}}}}}}+\ensuremath{\frac{f_{1}(S_{i})}{\mbox{\ensuremath{\mbox{\ensuremath{X_{0}(S_{i})}}-1}}}}+\ensuremath{\frac{f_{1}^{2}(S_{i})}{\ensuremath{\mbox{\ensuremath{X_{0}(S_{i})\left[\ensuremath{\mbox{\ensuremath{X_{0}(S_{i})}}}-1\right]}}}}}}\right]}} & (S_{i}\equiv S_{j})
\end{cases}\label{eq:5}
\end{equation}
where $R_{0}()$ and $X_{0}()$ are the equivalent old reaction rate
and old population. Therefore, the reaction rate updates depend only
on values of the $f_{1}$ parameters and the old populations of the
clusters.

If both $f_{1}$ and $f_{2}$ are zero, the cluster $S_{i}$ is not
affected by the selected event, but we need to check whether $S_{i}$
can engage in any $2^{\text{nd}}$-order reactions with those clusters
$S_{j}$ that are affected by the recent reaction event. These clusters
are stored in the dynamic array mentioned previously. The last step
of this process is to check whether the clusters in the dynamic array can
form $2^{\text{nd}}$-order reactions with one another. Some of these
reactions, which may have been skipped in previous steps because of the way
the reaction \emph{keys} are assigned, are now accounted for in this
step. If any pair of defect-clusters $S_{i}$ and $S_{j}$ can react,
the corresponding reactions --as well as the total reaction rate--
can be updated accordingly using eqs.\ \ref{eq:3}, \ref{eq:4}, and \ref{eq:5}.
\end{doublespace}


\begin{doublespace}

\subsection{Implementation of the $\mathbf{\tau}$-leaping method in SCD}\label{implement}
\end{doublespace}

\begin{doublespace}
In this section, we describe our implementation of the $\tau$-leaping method
within SCD. The method has been previously implemented on top of the
direct SSA algorithm \cite{tauleap}. However, implementation of $\tau$-leaping
in an open system where new species are constantly added to or removed
from the reaction network, as is the case of SCD simulations of irradiated
materials, has not been attempted to our knowledge.  Employing hash
tables, we now show how $\tau$-leaping can be added to the SCD method
to make the simulations more efficient. Several improvements have
been proposed to the $\tau$-leaping method since it was first proposed
by Gillespie \cite{tauleap,gillespie2003,chat2005,cao2006,tian2006,cao2005}, including efficient simulations of
stiff reaction networks \cite{hasel2002,rath2003,samant2005} or prevention of meaningless negative
species populations that can be caused by leaping \cite{cao2005b}. Cao et
al developed an efficient $\tau$-leaping SSA algorithm that avoids
having to solve a complicated set of partial differential equations
suggested in Ref.~\cite{cao2006}. For the sake of clarity and to better
explain our implementation of $\tau$-leaping in the SCD algorithm,
here we briefly summarize Cao {\it et al.}'s algorithm. The reader is referred
to the original paper for more details of the method and the underlying
theory \cite{cao2006}.

In Cao {\it et al.}'s approach, the set of all existing reactions is divided into
two non-overlapping subsets: the critical subset includes all reactions
that are within $n_{cr}$ (a pre-defined integer) firings away from
extinguishing one of the component reactants and the noncritical subset
includes all the other reactions. We add all $0^{\text{th}}$-order
defect insertion reactions to the critical subset in which every reaction
is advanced one at a time, just like in the \emph{direct} SSA method.
To enable efficient $\tau$-leaping over the noncritical reaction
subset, we make use of two more hash tables. The first one $\overrightarrow{P}$
is used to store the noncritical species, each element in $\overrightarrow{P}$
containing the species' attributes such as its \emph{key}, population
count and several additional parameters $g$, $\mu$, $\sigma^{2}$
and $O$'s as defined below. The second hash table $\overrightarrow{\mathbb{J}}$
contains the noncritical reactions. Similar to the regular reaction
hash, each element of $\mathbb{\overrightarrow{\mathbb{J}}}$ contains
the key and the rate of a noncritical reaction. Denoting the lower
number of clusters among the two reactants as $X_{min}$, a safe leap
time $\tau$ for every noncritical reaction in $\overrightarrow{\mathbb{J}}$
is selected as

\begin{equation}
\tau^{\prime}=\underset{P_{i}\in\overrightarrow{P}}{\min}\left\{ \dfrac{\max\{\epsilon X(P_{i})/g(P_{i}),1\}}{\left|\mu(P_{i})\right|},\dfrac{\max\{\epsilon X(P_{i})/g(P_{i}),1\}^{2}}{\sigma^{2}(P_{i})}\right\} \label{eq:6}
\end{equation}

with 

\begin{equation}
\mu(P_{i})=\sum_{\mathbb{J}_{j}\in\overrightarrow{\mathbb{J}}}\nu_{ij}J_{j}\label{eq:7}
\end{equation}

\begin{equation}
\sigma^{2}(P_{i})=\sum_{\mathbb{J}_{j}\in\overrightarrow{\mathbb{J}}}\nu_{ij}^{2}J_{j}\label{eq:8}
\end{equation}

The value of $g_{i}$ depends on the highest order $O_{i}$ of any
reaction in which the noncritical cluster $P_{i}$ appears as a reactant.
As appropriate for our model of irradiated materials, we categorize
these reaction-order parameters into three different types: $1^{\text{st}}$-order
($O_{1}$), $2^{\text{nd}}$-order ($O{}_{2}$), and $2^{\text{nd}}$-order
with like reactants ($O{}_{3}$). When a reaction becomes critical
or no longer exists due to exhaustion of one or both of its reactants,
$O_{i}$ parameters of the participating reactants are updated accordingly.
The values of $\mu$ and $\sigma^{2}$ for each $P_{i}$ are also
updated every time a reaction involving a noncritical cluster is analyzed.

To determine the value of the leap time $\tau$, our algorithm inspects
all clusters $P_{i}$ stored in the noncritical-reactant hash $\overrightarrow{P}$,
determines the highest order of their associated reactions and calculates
the corresponding values of $g_{i}$. When $O{}_{2}(P_{i})=0$, the
highest order of reactions involving species $P_{i}$ is $1^{\text{st}}$-order,
and the corresponding value of $g(P_{i})$ is 1. A positive $O{}_{2}(P_{i})$
indicates that $P_{i}$ takes part in at least one $2^{\text{nd}}$-order
reaction in which case $g(P_{i})$ is taken to be 2. However when
a reaction exists that can involve two clusters of species $P_{i}$,
$O{}_{3}(P_{i})$ will also be positive and the value of $g(P_{i})$
is determined instead as 

\begin{equation}
g(P_{i})=\left[2+\dfrac{1}{X(P_{i})-1}\right]\label{eq:9}
\end{equation}

After a safe value of the leaping time is estimated as described above,
the number of times $k_{i}$ each reaction $\mathbf{\mathbb{J}}_{i}\in\overrightarrow{\mathbb{J}}$
in the noncritical reaction hash will fire during this interval is
computed as a Poisson random variable $\mathcal{P}(J_{i},\tau)$.
However the reactions are not executed immediately as it is still
necessary to ensure that none of the noncritical reactant populations
$\overrightarrow{P}$ becomes negative after $\tau$-leaping is performed
on all reactions in the current noncritical reaction hash$\overrightarrow{\mathbb{J}}$.
To ensure that all species populations remain non-negative after $\tau$-leaping,
the total number $k^{tot}(P_{i})$ of reaction events reducing the
population of species $P_{i}$ is obtained by summing $k_{j}$ over
all noncritical reactions $\mathbb{J}_{j}$ consuming $P_{i}$ during
the leap time $\tau$. Only when the population of every noncritical
cluster $X(P_{i})$ is found to be larger than $k^{tot}(P_{i})$,
every reaction $\mathbb{J}_{i}$ stored in the noncritical reaction
hash is executed $k_{i}$ times and the $f_{1}$ and $f_{2}$ parameters
of reactant clusters $S_{1}$ or $S_{2}$ are updated accordingly;
otherwise, the value of $\tau$ is reduced, new firing times $k_{i}$'s
are determined and the previous non-negativity condition is re-examined.
Should the need for reduction in $\tau$ persist, $\tau$-leaping
is abandoned in favor of the \emph{direct} (single-reaction) SSA algorithm
for some number of SSA steps (200 steps in simulations described in
Section \ref{numero}) after which $\tau$-leaping is resumed.
\end{doublespace}

\begin{doublespace}

\subsection{Controlling the simulation complexity using volume rescaling}
\end{doublespace}

\begin{doublespace}
The computational complexity of SCD simulations is largely defined
by the number of distinct cluster species currently present in the
defect population. This number can be controlled by the size of the
simulation volume. In selecting the volume, one needs to balance two
conflicting requirements: (1) defect cluster populations should be
statistically representative (which favors larger volumes) and (2)
the computational cost of SCD simulations should remain acceptable
(which favors smaller volumes). Here we introduce a method to balance
these two requirements through volume rescaling. 

Typically, at the start of a SCD simulation, most of defect clusters
are mobile and their volume concentrations as well as the concentration
and the net strength of pre-existing defect sinks (dislocations, grain
boundaries, etc) are low. Under such conditions, mobile defects diffuse
over long distances through the reaction volume before they meet a
reaction partner. However, as time proceeds and progressively more
defects are inserted by continued irradiation, clusters become more
numerous while smaller mobile clusters combine and form increasingly
larger clusters. Such kinetics result in a more or less steady reduction
in the lifetime and diffusion length of mobile clusters defined as
the average time and distance travelled by a mobile cluster from birth
to death, respectively. In a given defect population, the average
lifetime of a mobile cluster of species $S_{i}$ is the inverse of
the total rate of loss:
\begin{equation}
\mathcal{L}(S_{i})=D(S_{i})\sum_{l}Z_{il}\rho_{l}+\epsilon_{m}+\sum_{j}k_{ij}\dfrac{X(S_{j})}{V},\label{eq:11}
\end{equation}
while the maximum diffusion length among all mobile cluster species
$\overrightarrow{S^{\prime}}_{m}$ can be estimated as
\begin{equation}
l_{max}=\underset{S_{i}\in\overrightarrow{S^{\prime}}_{m}}{\max}l_{S_{i}},~~~l_{S_{i}}=\sqrt{\dfrac{D(S_{i})}{\mathcal{L}(S_{i})}} \label{eq:10}
\end{equation}
where $\mathcal{L}(S_{i})$ is the net rate of loss and $D(S_{i})$ is
the diffusion coefficient of the mobile cluster species $S_{i}$,
$Z_{il}$ is the strength of a given sink of type $l$ with respect
to the same species (the sink's ability to remove clusters $S_{i}$),
$\rho_{l}$ is the volume density of sinks of type $l$, $\epsilon_{m}$
is the total rate of all dissociation reactions leading to splitting
clusters $S_{i}$, and $k_{ij}$ is the reaction rate of a $2^{\text{nd}}$-order
reaction between the mobile defect $S_{i}$ and the defect cluster
of type $S_{j}$ with a population of $X(S_{j})$. Here $\overrightarrow{S^{\prime}}_{m}$
denotes the set of all mobile species that will possibly appear in
the simulation volume, not limited to only those exist at the current
time-step.

The significance of parameter $l_{max}$ is that it defines the range
of distances beyond which neighboring reaction sub-volumes are no
longer exchanging their reactants (defect clusters). Thus, reaction
volumes with linear dimensions exceeding $l_{max}$ can be viewed
as causally isolated from each other. Typically, as a SCD simulation
progresses $l_{max}$ decreases due to a more or less steady increase
in the magnitude of the last term on the right hand side of Eq.\ref{eq:11}.
A significant reduction in $l_{max}$ justifies an appropriate reduction
in the reaction volume, $V_{new}=\gamma V_{old}\geq l_{max}^3$ (with $\gamma<1$). The
essence of our volume rescaling method is that when conditions for
volume reduction conditions are satisfied, the cluster population
is reduced by allowing every cluster to be randomly eliminated with
probability $(1-\gamma)$ before resuming the SCD simulation. Such a
volume reduction procedure allows to maintain the size of the reaction
network approximately constant even when damage accumulation increases
the volume density of defects by orders of magnitude. However, volume
rescaling should be avoided when there are large fluctuations in the
defect population, for example right after a massive defect insertion
event.
\end{doublespace}

\subsection{Algorithm implementation}

\begin{doublespace}
In this section we present the key algorithmic
elements of our improved SCD method in pseudocode format, including construction of hash
tables for noncritical reactions and defect clusters and an algorithm
for estimating a safe leap time $\tau$ in SCD. In the following $R(S_{1},S_{1})$
is the rate of a binary reaction\textbf{ }$\mathbf{\mathbb{R}}(S_{1},S_{2})$
between species $S_{1}$ and $S_{2}$. Similarly, $J(P_{1},P_{2})$
denotes the rate of the noncritical reaction $\mathbb{J}(P_{1},P_{2})$
between two noncritical species $P_{1}$ and $P_{2}$. The set of all critical reactions is represented by $\overrightarrow{\mathbb{R}}_{cr}$. For a $1^{\text{st}}$-order
reaction \textbf{$\mathbb{R}(S_{1})$ }or $\mathbb{J}(S_{1})$, $S_{1}$
represents its one and only reactant cluster. $X(S_{i})$ denotes
the population (number of units) of cluster species $S_{i}$ in the
reaction volume.

\subsubsection{Construction of the noncritical hash tables }\label{alg0}
\end{doublespace}

\noindent \begin{flushleft}
\doublebox{\begin{minipage}[t]{1\columnwidth}%
\begin{enumerate}
\item \noindent If $\mathbb{R}$ is a $2^{\text{nd}}$-order reaction:

\begin{enumerate}
\item \noindent If $S_{1}\equiv S_{2}$ (reaction between two like clusters):

\begin{enumerate}
\item \noindent If $X(S_{1})>=n_{cr}+2$ (reaction $\mathbb{R}$ is noncritical,
and cluster $S_{1}$ is noncritical, i.e. $\mathbb{R}(S_{1})\equiv\mathbb{J}(S_{1})\in\overrightarrow{\mathbb{J}}$
and $S_{1}\equiv P_{1}\in\overrightarrow{P}$)

\begin{itemize}
\item \noindent If $\mathbb{J}(P_{1})$ does not exist: add $\mathbb{J}(P_{1})$
into the noncritical reaction hash $\overrightarrow{\mathbb{J}}$
and update $O{}_{2,3}(P_{1})\leftarrow O{}_{2,3}(P_{1})+1$.
\item \noindent Else: update $J(P_{1})\leftarrow R(S_{1})$ and $X_{min}(\mathbb{J})\leftarrow X(S_{1})$,
reset $k\left[\mathbb{J}(P_{1})\right]\leftarrow0$.
\item \noindent Locate $P_{1}$ in the noncritical cluster hash $\overrightarrow{P}$

\begin{itemize}
\item \noindent If $P_{1}$ does not exist: add $P_{1}$ to $\overrightarrow{P}$
with $\mu_{tot}(P_{1})\leftarrow\mu\left[J(P_{1})\right]$ and $\sigma_{tot}^{2}(P_{1})\leftarrow\sigma^{2}\left[J(P_{1})\right]$
as determined by eqs.\ \ref{eq:7} and \ref{eq:8}.
\item \noindent Else: update $\mu_{tot}(P_{1})\leftarrow\mu_{tot}(P_{1})+\left\{ \mu\left[J(P_{1})\right]\right\} _{new}-\left\{ \mu\left[J(P_{1})\right]\right\} _{old}$
and $\sigma_{tot}^{2}(P_{1})\leftarrow\sigma_{tot}^{2}(P_{1})+\left\{ \sigma^{2}\left[J(P_{1})\right]\right\} _{new}-\left\{ \sigma^{2}\left[J(P_{1})\right]\right\} _{old}$.
\end{itemize}
\end{itemize}
\item \noindent Else (\textbf{$\mathbb{R}$} is critical): 

\begin{itemize}
\item \noindent If $\mathbb{P}(P_{1})$ exists: locate $P_{1}$ in $\overrightarrow{P}$.
If $P_{1}$ exists: update $O{}_{2,3}(P_{1})\leftarrow O{}_{2,3}(P_{1})-1$,
$\mu_{tot}(P_{1})\leftarrow\mu_{tot}(P_{1})-\mu\left[J(P_{1})\right]$,
$\sigma_{tot}^{2}(P_{1})\leftarrow\sigma_{tot}^{2}(P_{1})-\sigma^{2}\left[J(P_{1})\right]$.
If $O{}_{1}(P_{1})=O{}_{2}(P_{1})=O{}_{3}(P_{1})=0$, remove $P_{1}$
from the noncritical cluster hash $\overrightarrow{P}$.
\item \noindent Remove $\mathbb{J}(P_{1})$ from the noncritical reaction
hash $\overrightarrow{\mathbb{J}}$.
\end{itemize}
\end{enumerate}
\item \noindent Else (reaction between unlike species $S_{1}\neq S_{2}$
): 

\begin{enumerate}
\item \noindent If $min\left\{ X(S_{1}),X(S_{2})\right\} >n_{cr}$ (reaction
$\mathbb{R}$ is noncritical, and clusters $S_{1}$, $S_{2}$ are
noncritical, i.e. $\mathbb{R}(S_{1},S_{2})\equiv\mathbb{J}(S_{1},S_{2})\in\overrightarrow{\mathbb{J}}$
and $S_{1,2}\equiv P_{1,2}\in\overrightarrow{P}$)

\begin{itemize}
\item \noindent If $\mathbb{J}(P_{1},P_{2})$ does not exist: add $\mathbb{J}(S_{1},S_{2})$
into the noncritical reaction hash $\overrightarrow{\mathbb{J}}$,
update $O{}_{2}(P_{1,2})\leftarrow O{}_{2}(P_{1,2})+1$.
\item \noindent Else: update $J(P_{1},P_{2})\leftarrow R(S_{1},S_{2})$
and $X_{min}(\mathbb{J})\leftarrow min\left\{ X(S_{1}),X(S_{2})\right\} $,
reset $k\left[\mathbb{J}(P_{1},P_{2})\right]\leftarrow0$.
\item \noindent Locate $P_{1}$ and $P_{2}$ in the noncritical cluster
hash $\overrightarrow{P}$

\begin{itemize}
\item \noindent If $P_{1}$ and/or $P_{2}$ do not exist: add them into
$\overrightarrow{P}$ with $\mu_{tot}(P_{1,2})\leftarrow\mu\left[J(P_{1},P_{2})\right]$
and $\sigma_{tot}^{2}(P_{1,2})\leftarrow\sigma^{2}\left[J(P_{1},P_{2})\right]$
as determined by eqs.\ \ref{eq:7} and \ref{eq:8}.
\item \noindent Else: update $\mu_{tot}(P_{1,2})\leftarrow\mu_{tot}(P_{1,2})+\left\{ \mu\left[J(P_{1},P_{2})\right]\right\} _{new}-\left\{ \mu\left[J(P_{1},P_{2})\right]\right\} _{old}$
and $\sigma_{tot}^{2}(P_{1,2})\leftarrow\sigma_{tot}^{2}(P_{1,2})+\left\{ \sigma^{2}\left[J(P_{1},P_{2})\right]\right\} _{new}-\left\{ \sigma^{2}\left[J(P_{1},P_{2})\right]\right\} _{old}$. 
\end{itemize}
\end{itemize}
\item \noindent Else: (\textbf{$\mathbb{R}$} is critical):

\begin{itemize}
\item \noindent If $\mathbb{J}(P_{1},P_{2})$ exists: locate $P_{1}$ and
$P_{2}$ in $\overrightarrow{P}$. If $P_{1}$ or $P_{2}$ exists:
update $O{}_{2}(P_{1,2})\leftarrow O{}_{2}(P_{1,2})-1$, $\mu_{tot}(P_{1,2})\leftarrow\mu_{tot}(P_{1,2})-\mu\left[J(P_{1},P_{2})\right]$,
$\sigma_{tot}^{2}(P_{1,2})\leftarrow\sigma_{tot}^{2}(P_{1,2})-\sigma^{2}\left[J(P_{1},P_{2})\right]$.
If $O{}_{1}(P_{1,2})=O{}_{2}(P_{1,2})=O{}_{3}(P_{1,2})=0$, remove
$P_{1}$ and/or $P_{2}$ from the noncritical cluster hash $\overrightarrow{P}$.
\item \noindent Remove $\mathbb{J}(P_{1},P_{2})$ from the noncritical reaction
hash $\overrightarrow{\mathbb{J}}$.
\end{itemize}
\end{enumerate}
\end{enumerate}
\item \noindent Else: $\mathbb{R}$ is a $1^{\text{st}}$-order reaction
(emission or absorption of a defect cluster at sinks). Follow similar
steps as described in 1(a), except that the noncritical
condition for cluster $S_{1}$ is $X(S_{1})>n_{cr}$ in this case.\end{enumerate}

\end{minipage}}
\par\end{flushleft}
\subsubsection{Reaction update loop}\label{alg1}
\noindent \begin{flushleft}
\doublebox{\begin{minipage}[t]{1\columnwidth}%
\begin{enumerate}
\item Remove all illegal reactions whose reactants are no longer exist from the reaction hash $\overrightarrow{\mathbb{R}}$. Starting from
the first cluster $S_{1}$ in the all-cluster hash $\overrightarrow{S}_{all}$:
\item If $f_{1}(S_{1})\text{ \ensuremath{\neq} }0$ or $f_{2}(S_{1})\text{ \ensuremath{\neq} }0$
($S_{1}$ is affected by the recently executed event):

\begin{enumerate}
\item If $f_{1}(S_{1})\text{ \ensuremath{\neq} }0$ and $f_{2}(S_{1})=0$
($S_{1}$ is only the reactant not the product of the previously executed
event):

\begin{enumerate}
\item Check if $S_{1}$ is immobile, skip to ii; else: find $1^{\text{st}}$-order
reaction channels associated with $S_{1}$ and update the reaction
rates using Eq. 5.
\item Loop through the mobile-cluster hash $\overrightarrow{S}_{m}$, determine
if $key(S_{1})\geq key(S_{2})$ where $S_{2}$ denotes the mobile
cluster and find the associated reaction channel $\mathbb{R}(S_{1},S_{2})$. 

\begin{enumerate}
\item If $\mathbb{R}(S_{1},S_{2})$ exists: update values of $\mathbb{R}(S_{1},S_{2})$ and the total
rate $R_{tot}$ using Eqs. 6 and 7.
\item Else: calculate the reaction rate \textbf{$R(S_{1},S_{2})$} between
clusters $S_{1}$ and $S_{2}$, add $\mathbb{R}(S_{1},S_{2})$ to
the reaction hash \textbf{$\overrightarrow{\mathbb{R}}$} and update
the total reaction rate, $R_{tot}\leftarrow R_{tot}+R(S_{1},S_{2})$. 
\end{enumerate}
\end{enumerate}
\item Else if $f_{2}(S_{1})\neq0$ (the cluster has just been created):
similar to 2(a), except that all reactions associated with $S_{1}$
will be added directly into the reaction hash $\overrightarrow{\mathbb{R}}$.
It is not necessary to locate these reactions in $\overrightarrow{\mathbb{R}}$
since they are completely new reactions.
\end{enumerate}
\item Else $f_{1}(S_{1})=f_{2}(S_{1})=0$ (the cluster does not participate
in the previous reaction): 

\begin{enumerate}
\item Loop through the effected clusters $S_{2}$ contained in the dynamic
array and evaluate these following conditions: 1) $key(S_{1})>key(S_{2})$
and $S_{1}$ is mobile, 2) $S_{1}$ is immobile while $S_{2}$ is
mobile.
\item If any of those conditions is satisfied: find the associated reaction
$\mathbf{\mathbb{R}}(S_{1}^{\prime},S_{2}^{\prime})$ ($S_{1}^{\prime}$
is the larger value of $S_{1}$ and $S_{2}$, the other is $S_{2}^{\prime}$). 

\begin{enumerate}
\item If $\mathbf{\mathbb{R}}(S_{1}^{\prime},S_{2}^{\prime})$ exists: update
$R(S_{1}^{\prime},S_{2}^{\prime})$ and $R_{tot}$ using Eq. 6 and
7.
\item Else: calculate the reaction rate $R(S_{1}^{\prime},S_{2}^{\prime})$
between clusters $S_{1}$ and $S_{2}$, add $\mathbf{\mathbb{R}}(S_{1}^{\prime},S_{2}^{\prime})$
into the reaction hash $\overrightarrow{\mathbb{R}}$ and update the total reaction
rate, $R_{tot}\leftarrow R_{tot}+R(S_{1}^{\prime},S_{2}^{\prime})$
.
\end{enumerate}
\end{enumerate}
\item Proceed to the next cluster in the all-cluster hash $\overrightarrow{S}_{all}$
and repeat Step 2 until reaching the last cluster.
\item Loop through the clusters contained in the dynamic array. For all
possible pairs of $(S_{1},S_{2})$, if at least one of the clusters
in the pair is mobile: find the associated reaction $\mathbb{R}(S_{1}^{\prime},S_{2}^{\prime})$
($S_{1}^{\prime}$ is the larger value of $S_{1}$ and $S_{2}$, the
other is $S_{2}^{\prime}$). 

\begin{enumerate}
\item If $\mathbb{R}(S_{1}^{\prime},S_{2}^{\prime})$ exists: update $R(S_{1}^{\prime},S_{2}^{\prime})$
and $R_{tot}$ using eqs.~\ref{eq:4} and \ref{eq:5}.
\item Else: calculate the reaction rate $R(S_{1}^{\prime},S_{2}^{\prime})$
between clusters $S_{1}$ and $S_{2}$, add $\mathbb{R}(S_{1}^{\prime},S_{2}^{\prime})$
into the reaction hash $\overrightarrow{\mathbb{R}}$ and update the total
reaction rate, $R_{tot}\leftarrow R_{tot}+R(S_{1}^{\prime},S_{2}^{\prime})$.
\end{enumerate}
\item Locate in the all-cluster hash $\overrightarrow{S}_{all}$ the same
clusters $S_{i}$ that are stored in the dynamic array and reset the
values of $f_{1}$ and $f_{2}$: $0\leftarrow f_{1}(S_{i})$ and $0\leftarrow f_{2}(S_{i})$
and clear the dynamic array.
\end{enumerate}
If the $\tau$-leaping method is implemented, update the noncritical
cluster hash $\overrightarrow{P}$ and noncritical reaction hash $\overrightarrow{\mathbb{J}}$
at the end of Steps 2, 3 and 5 above as described in Algorithm \ref{alg0}.%
\end{minipage}}
\par\end{flushleft}

\begin{doublespace}

\subsubsection{Main event loop}
\end{doublespace}

\noindent \begin{flushleft}
\doublebox{\begin{minipage}[t]{1\columnwidth}%
\begin{enumerate}
\item If the simulation is resumed from a pre-existing one, enter input
data into the hash tables $\overrightarrow{S}_{all}$, $\overrightarrow{S}_{m}$,
and $\overrightarrow{\mathbb{R}}$. Set the appropriate initial time
and compute the total rate $R_{tot}$ of all reactions associated
with existing defect clusters in $\overrightarrow{S}_{all}$ . Skip
to Step 3.
\item Update the $\overrightarrow{S}_{all}$ , $\overrightarrow{S}_{m}$
and $\overrightarrow{\mathbb{R}}$ hashes and the total reaction rate $R_{tot}$
as described in Algorithm \ref{alg1} Perform volume rescaling if the conditions
in eq.~\ref{eq:11} are satisfied.
\item If SSA has run less than $N_{SSA}$ steps: select and execute a reaction
event $\mathbb{R}(S_{1},S_{2})\in\overrightarrow{\mathbb{R}}$ and
calculate the time to next reaction event using eqs. \ref{eq:dt} and \ref{eq:random variate}, store
identities of the effected clusters in the dynamic array and return
to Step 2 until the final time is reached; else: go to Step 4.
\item Reset $N_{SSA}\leftarrow0$. If the noncritical reaction hash $\overrightarrow{\mathbb{J}}$
is empty: $\tau$-leaping cannot be performed, return to Step 3; else:

\begin{enumerate}
\item Calculate the value of $g(P_{i})$ for each cluster $P_{i}\in\overrightarrow{P}$
based on the values of its $O_{i}$ parameters as described in Section \ref{implement}.
\item Determine the value of the noncritical time leap $\tau^{\prime}$
using eq.\ \ref{eq:6}.
\item Calculate the total reaction rate $R_{cr}$ of all the critical reactions in $\overrightarrow{\mathbb{R}}_{cr}$ and the critical time leap $\tau^{\prime\prime}$ using eq.\ \ref{eq:dt}.
\item If $\tau^{\prime}$ is less than some small $n$-multiple (we set
$n$ equal 10) of $1/R_{tot}$, temporarily abandon $\tau$-leaping and
return to Step 3.
\item Else:

\begin{enumerate}
\item Take the leap time to be the smaller value of $\tau^{\prime}$ and
$\tau^{\prime\prime}$, $\tau=min\left\{ \tau^{\prime},\tau^{\prime\prime}\right\} $.
\item Calculate the number of times each reaction $\mathbf{\mathbb{J}}_{i}\in\overrightarrow{\mathbb{J}}$
will fire during this time interval $[t,t+\tau$) as described in
Section \ref{implement}.
\item If $\mathcal{P}\left(J\tau\right)>X(P_{1})$ (if $\mathbb{J}_{i}$
is a $1^{\text{st}}$-order reaction$ $) or $\mathcal{P}\left(J_{i}\tau\right)>min\left\{ X(P_{1}),X(P_{2})\right\} $
(if $\mathbb{J}_{i}$ is a $2^{\text{nd}}$-order reaction): reduce
$\tau^{\prime}$ by half and return to Step 4(d). Else: assign $k(\mathbb{J}_{i})\leftarrow\mathcal{P}\left(J_{i}\tau\right)$,
$k(P_{1,2})\leftarrow k(P_{1,2})+k\left[\mathbb{J}(P_{1},P_{2})\right]$
. If $k(P_{1,2})>X(P_{1,2})$: reduce $\tau^{\prime}$ by half and
return to Step 4(d).
\item Execute $\mathbb{J}_{i}\in\overrightarrow{\mathbb{J}}$ a number of
$k(\mathbf{\mathbb{J}}_{i})$ times. Store the identities of the effected
clusters in the dynamic array if $k(\mathbf{\mathbb{J}}_{i})>0$.
Update $t\leftarrow t+\tau$, then return to Step 2 or stop if the
final time has been reached.
\item If $\tau^{\prime\prime}\leq\mbox{\ensuremath{\tau}}^{\prime}$: select
and execute a critical reaction event $\mathbf{\mathbb{R}}(S_{1},S_{2})\in\overrightarrow{\mathbb{R}}_{cr}$
and store identities of the effected clusters in the dynamic array.
If an insertion event is selected, process the event and store the
identities of the new clusters in the the dynamic array, then return
to Step 2 or stop the simulation if the final time has been reached.\end{enumerate}
\end{enumerate}
\end{enumerate}
\end{minipage}}
\par\end{flushleft}

\begin{doublespace}

\section{Triple ion-beam irradiation of bcc-Fe thin film}\label{numero}
\end{doublespace}

\subsection{Simulation}

\begin{doublespace}
Materials performance in nuclear fusion reactors is expected to degrade
as a consequence of prolonged exposure to neutron irradiation. However,
neutron irradiation experiments are costly, irradiation facilities
are scarce and presently achievable neutron fluxes are low requiring
years of exposure before material specimens receive a significant
dose of irradiation. As a faster and more cost effective alternative
for assessing irradiation-induced changes in physical and mechanical
properties of materials, ion beam experiments are used for accelerated
testing of material degradation because ion cascades can produce damage
similar to neutron irradiation but on a much shorter time scale. In
addition to the displacement damage, material exposure to fast neutrons
results in simultaneous formation of He and H atoms through nuclear
transmutation reactions.  To mimic such specific conditions properly,
triple ion beam irradiation can be used in which ions of He and H
are co-implanted, either sequentially or concurrently, with the heavy
ions imparting the primary (displacement) damage. Recent triple-beam
experiments of this kind conducted on iron crystals have revealed
pronounced synergistic effects associated with co-implantation of
He and H under irradiation by self-ions of Fe \cite{tanaka2004}. Specifically, the amount of measurable swelling increased several
fold when all three ion species were implanted simultaneously, relative
to baseline sequential dual $\text{Fe}^{3+}/\text{ He}^{+}$ and $\text{Fe}^{3+}/\text{ H}^{+}$
irradiations.

As previously discussed, ODE-based simulations methods have so far
proven incapable of coping with complex cluster species with more
than two size attributes, as is the case of triple-beam irradiations
reported in Ref.~\cite{tanaka2004}. Here we show that our enhanced SCD method
is capable of simulating of complex defect microstructures in pure
iron subjected to simultaneous irradiation with $\text{Fe}^{3+}$
ions and co-implantation with $\text{He}^{+}$ and $\text{H}^{+}$
ions. In setting up our model and SCD simulations we mimic as close
as possible irradiation conditions used in the triple ion-beam experiments
performed by Tanaka and coworkers. The model parameters used in SCD
simulations reported here are the same as in Ref.~\cite{marian2011}. 

\begin{figure}[h]
\noindent \begin{centering}
\includegraphics[width=1.0\textwidth]{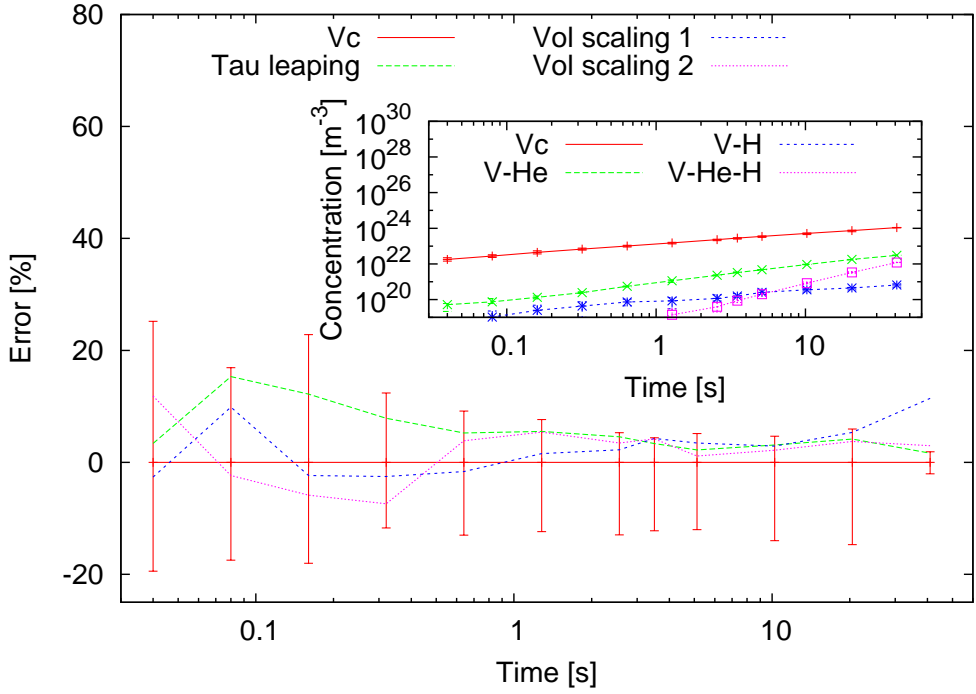}
\par\end{centering}
\caption{Statistical errors of various enhancement methods compared to the
original SCD model. The specimen is under triple ion irradiation of
$\text{F\ensuremath{e^{3+}}}$, $\text{H\ensuremath{e^{+}}}$ and
$\text{H}^{+}$, total irradiation time is 40.96 seconds and the temperature
is 783K. The inset shows the concentrations of various defect-cluster
types as functions of irradiation time, in this case the simulation
is carried out using the original SCD algorithm with no improvement.
Vol scaling 1 uses $\gamma=0.9999,$ and Vol scaling 2 uses $\gamma=0.99999$.}\label{jop}
\end{figure}

We have performed simulations of triple-beam
irradiation and tracked the accumulation of pure vacancy (V), V-He,
V-H, and V-He-H clusters in the simulation volume using first direct
(exact) SCD simulations \cite{marian2011} and then repeated the same simulations
after turning on, one by one, the various enhancements described in
the preceding sections. The inset to Figure \ref{jop} shows the concentrations
of various types of clusters as functions of simulated irradiation
time as obtained with the original (unenhanced) SCD method. 
Each curve
in the inset was obtained by averaging over five independent simulations
starting from different random seeds. The main figure shows the relative
deviation from the reference (unenhanced) simulations in the net
vacancy cluster population obtained in SCD simulations with enhancements.
For consistency, five independent simulations were performed for every
enhancement. The error bars shown on the plot can be used as a measure
of statistical significance of the observed deviations. As the figure
shows, the results of enhanced SCD simulations fall within the statistical
errors to the exact (reference) simulations which verifies that the
approximations used here to improve computational efficiency of SCD
simulations, namely $\tau$-leaping and volume rescaling, are not
distorting the simulated kinetics of damage accumulation (for simulations
shown in Figure 1 we used the following values of runtime parameters:
$n_{cr}=10$ and $\epsilon=0.03$ for $\tau$-leaping and $\gamma$
ranging from 0.99999 to 0.9999 for volume rescaling). The ratio $\gamma$
can be reduced further to achieve even greater speedup, but accuracy
is what we prefer here since we have already managed to reduce the computing
time significantly with the current simulations.
\end{doublespace}

\begin{doublespace}

\subsection{Performance}
\end{doublespace}

\begin{doublespace}
First of all, a rather significant --a factor of 20 or higher-- speedup
in SCD simulations is attained simply due to a greater efficiency
of the incremental updates of the evolving reaction network and associated
reaction rates, as described in Sections \ref{dynnetwork} and \ref{update}. This is a general
improvement resulting from a better implementation of the standard
SCD algorithm reported in Ref.~\cite{marian2011}. We use the efficiency of our
standard SCD simulations with incremental updates as a reference comparison
with further enhancements. 

\begin{figure}[h]
\centering
\subfigure[]{
   \includegraphics[width=0.7\textwidth] {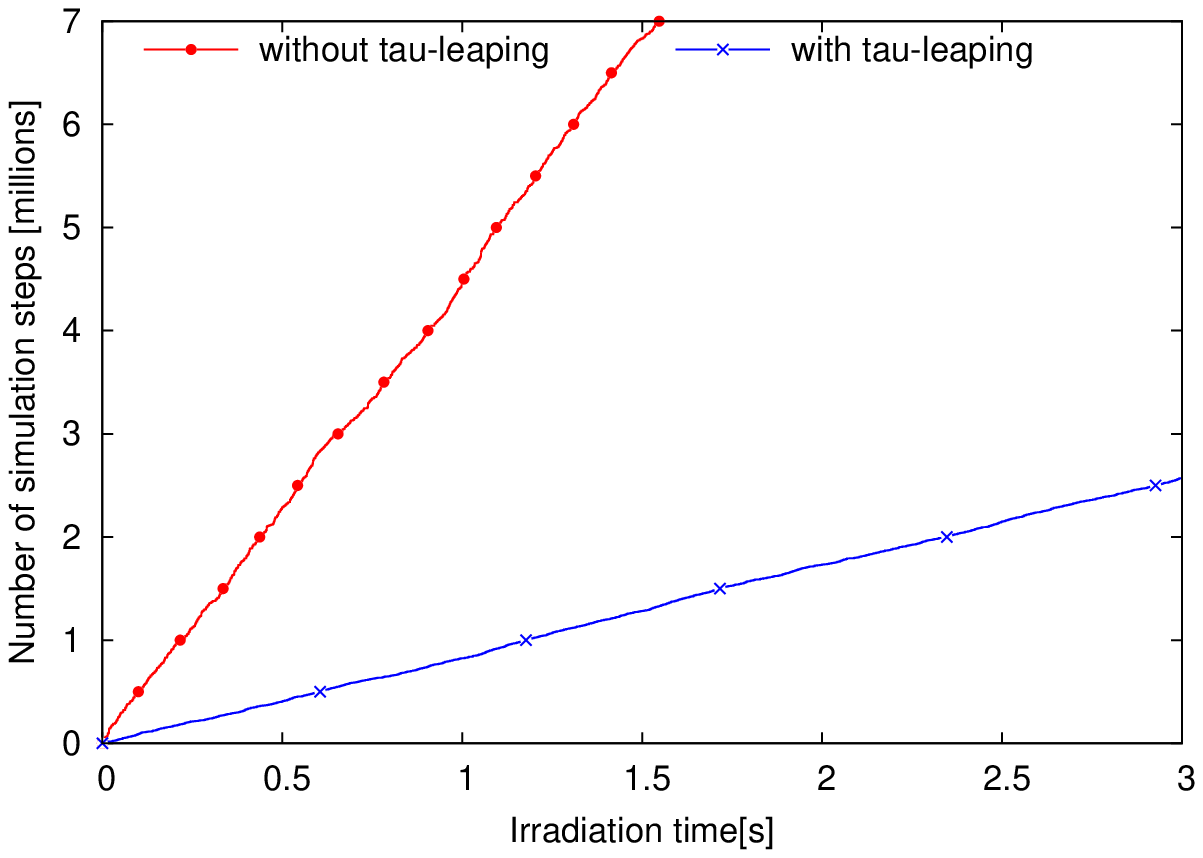}
   \label{subfig1}
 }
 \subfigure[]{
   \includegraphics[width=0.75\textwidth] {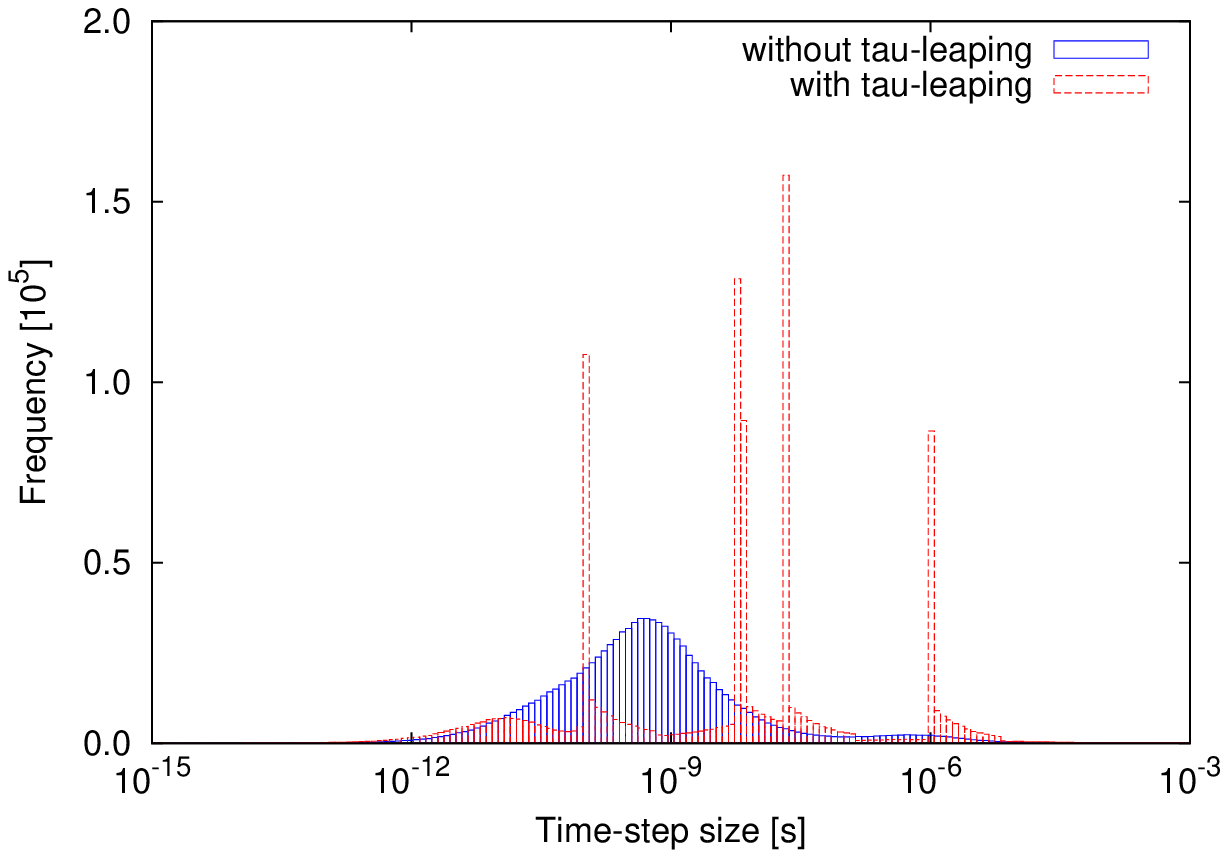}
   \label{subfig2}
 }
\caption{(a) Number of simulation steps as functions of the irradiation
time obtained from SCD simulations of triple-beam irradiation at 783K
with and without $\tau$-leaping implementation. (b) Distribution
of the time-steps in these two cases, here one millions time-steps
are collected and analyzed.}
\label{tauleap}
\end{figure}

We find that, in our SCD simulations of irradiated iron, conditions
for $\tau$-leaping are often satisfied and many reactions can be
allowed to fire at once rather than one at a time.  The key condition
for $\tau$-leaping to be accurate is that the change in the defect
population caused by a leaping step should not affect too much the
rates of existing reactions. Whenever it is safe to perform, $\tau$-leaping
results in longer time-steps compared to the standard (one reaction
at a time) SCD algorithm, as shown in Figure \ref{subfig1} for the same simulation
setup as described in the previous section. 
It is clear that, with $\tau$-leaping active, fewer short time-steps are taken than
in the direct SCD resulting in a total reduction in the number of
time-steps required to simulate the same evolution. Thus, the
total number of steps taken is reduced significantly, and so is the
overhead cost for system updating. This is confirmed in Figure \ref{subfig2},
where a histogram of the time-step size distribution for a $\tau$-leaping
simulation is plotted and compared to the same histogram obtained
from a standard SCD simulation. In addition to showing that the distribution
shifts to longer time-steps due to $\tau$-leaping, the same histogram
also shows that a few specific timesteps occur much more frequently
than the rest, and that they are clearly separated from each another.
The observed peaks in the distribution indicate that a handful of
noncritical reaction channels dominate the kinetics in our model,
and that our enhanced algorithm can identify and handle such reaction
channels efficiently with $\tau$-leaping. Specifically, the first
peak in Figure \ref{subfig2}, which ranges from 89.2 ps to 123 ps is dominated
by the absorption of SIA and SIA clusters by defect sinks. The second
group, whose reactions with time-steps between 5.01 ns and 25.1 ns
mostly consists of absorption of vacancies by sinks, and reactions
in the last group from 0.87 $\mu$s to 1.19 $\mu$s are predominantly
migration of vacancy clusters to defect sinks. As a result, $\tau$-leaping
is not only better for computational efficiency, but it also provides
very useful physical information by identifying the reactions that control
the kinetic evolution of the system. This has implications beyond
efficiency improvements because it can indicate where to focus the
efforts to calculate the physical parameters that matter the most with maximum accuracy.
This can potentially be helpful in uncertainty quantification of the models and/or to learn where to devote efforts to improve the physical parameterization.

\begin{figure}[h]
\centering
\includegraphics{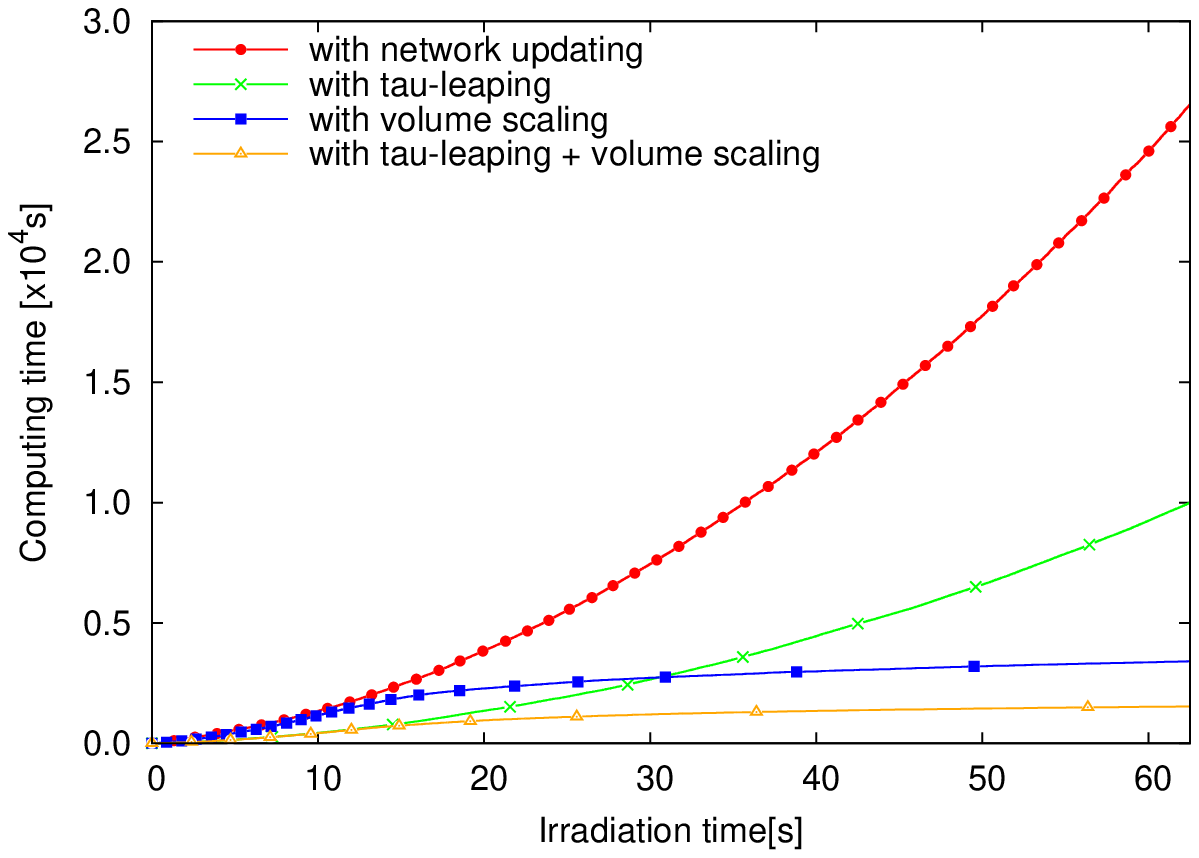}
\caption{Comparison of computational cost of the SCD model with different enhancement
methods, here a scaling ratio, $\gamma=0.99999$ is used whenever
the volume scaling method is implemented.}
\label{rescale}
\end{figure}

To quantify the speedup gained from the enhancements described in
this paper, the computational cost of SCD simulations performed with
and without the enhancements is plotted in Figure \ref{rescale} as a function
of the simulated time. 
As the figure shows, significant gains in simulation
efficiency are realized using $\tau$-leaping and volume rescaling.
$\tau$-leaping is typically most efficient at early stages of SCD
simulations but its associated speedup is subsequently negated by
an increasing computational cost of updates of the growing reaction
network. Under such circumstances volume rescaling is prescribed to
control the size of a growing defect population. When used together,
these two enhancements significantly reduce the wall clock time of
a SCD simulation without detectable sacrifice in its accuracy. As
an example, Figure \ref{rescale} shows that while it took the original SCD algorithm
more than two days to achieve a trivial irradiation dose of 0.1 dpa,
it only takes the enhanced algorithm only about 12 hours to reach
a technologically significant dose of 50 dpa.
\end{doublespace}

\begin{doublespace}

\section{Conclusions}\label{conc}
\end{doublespace}

\begin{doublespace}
We have presented a computationally efficient implementation of the SCD
algorithm originally devised as an adaptation of the well-known SSA
method to simulations of complex microstructure evolution in irradiated
materials. The key advantage of the SCD method is that, unlike the
traditional ODE-based rate theory approaches that notoriously suffer
from combinatorial explosion, SCD handles with ease multi-species
populations of arbitrary complexity. However, early applications of
the original SCD algorithm to irradiated materials exposed several
computational bottlenecks, e.g.~wide disparity in reaction rates and
stiff kinetics. Enhancements presented in this paper are introduced
to address some of the bottlenecks in order to achieve reactor-relevant
irradiation doses at a reasonable computational cost. Gains in computational
efficiency of SCD simulations are achieved through the following:
incremental updates of the evolving reaction network, $\tau$-leaping
permitting multiple reaction events to take place over a single simulation
step, and volume rescaling to control the size of defect population.
Further enhancements to the SCD algorithm reported here are being
considered, e.g.~a more robust method for SCD simulations of stiff
reaction networks with wide spectra of reaction rates and an adaptive
mechanism for deciding which method is best to use at each particular
stage of an SCD simulation to optimize the overall computational performance.
Efficient parallelization of the SCD algorithm is another interesting
venue for further research, e.g.~following replication strategies
recently proposed in the context of parallel kinetic Monte Carlo algorithms
\cite{martinez2011}. We note that, in addition to our SCD development borrowing
heavily from the SSA method ideas, algorithmic enhancements reported
here can be re-used in other simulation contexts where reaction-diffusion
processes with dynamic species populations are of interest, such as
in combustion science, cellular process simulation, or chemical kinetics. 
\section*{Acknowlegments}
We would like to thank Daryl Chrzan, Athanasios Arsenlis, Alexander Stukowski, David Cereceda, and Ninh Le for many helpful discussions. TLH acknowledges support from the Lawrence Scholar Program, the UC Berkeley Chancellor's Fellowship, and the Nuclear Regulatory Commission Research Fellowship.
This work was performed under the auspices of the U.S. Department of Energy by Lawrence Livermore National Laboratory under Contract DE-AC52-07NA27344.
\end{doublespace}


\begin{thebibliography}{10}
\bibitem{bull}
R. Bullough, B. L. Eyre and K. Krishan. Cascade Damage Effects on the Swelling of Irradiated Materials. Proc. R. Soc. Lond. A 346 (1975) 81-102

\bibitem{ghoniem}
N. M. Ghoniem and M. L. Takata. A rate theory of swelling induced by helium and displacement damage in fusion reactor structural materials. Journal of Nuclear Materials 105 (1982) 276292

\bibitem{mansur1996}
L. K. Mansur. The reaction rate theory of radiation effects. JOM 48 (1996) 28-32.

\bibitem{koiwa1974}
M. Koiwa. On the Validity of the Grouping Method ÐComments on ÒAnalysis of the Clustering Process of Supersaturated Lattice Vacancies. Journal of the Physics Society of Japan 37 (1974) 1532.

\bibitem{sands2013}
A. E. Sand, S. L. Dudarev and K. Nordlund. High-energy collision cascades in tungsten: Dislocation loops structure and clustering scaling laws. EPL, 103 (2013) 46003

\bibitem{marian2011} J. Marian, V. V. Bulatov. Stochastic cluster dynamics method for simulations of multispecies irradiation damage accumulation. J. Nucl. Mater. 415
(2012) 84-95 

\bibitem{gillespie1977} D. T. Gillespie. A general method for numerically
simulating the stochastic time evolution of coupled chemical reactions.
J. Comput. Phys. 22 (1976) 403\textendash{}434

\bibitem{asher} U. M. Ascher, L. R. Petzold. Computer Methods
for Ordinary Differential Equations and Differential Algebraic Equations.
Philadelphia: Soc. Ind. Appl. Math. (1998)

\bibitem{gillespie1992} D. T. Gillespie. A rigorous derivation of the
chemical master equation. Physica A 188 (1992) 404-425 

\bibitem{cao2004} Y. Cao, H. Li, L. R. Petzold. Efficient formulation
of the stochastic simulation algorithm for chemically reacting systems.
J. Chem. Phys. 121 (2004) 4059\textendash{}4067

\bibitem{cai2007}
X. Cai, Exact stochastic simulation of coupled chemical reactions with delays. J Chem Phys. 126 (2007) 124108.

\bibitem{ahn2008}
Tae-Hyuk Ahn, Yang Cao, Layne T. Watson. Stochastic Simulation Algorithms for Chemical Reactions. BIOCOMP-2008, 431-436.

\bibitem{cain} Cain: Stochastic Simulations for Chemical Kinetics
(http://cain.sourceforge.net), StochKit: a Stochastic Simulation Toolbox
for Biology (http://www.cs.ucsb.edu/\textasciitilde{}cse/index2.php?software.html)

\bibitem{tauleap} D. T. Gillespie. Approximate accelerated stochastic
simulation of chemically reacting systems. J. Chem. Phys. 115 (2001)
1716\textendash{}33 

\bibitem{gibson2000}
M. A. Gibson and J. Bruck. Efficient Exact Stochastic Simulation of Chemical Systems with Many Species and Many Channels. J. Phys. Chem. A 104 (2000) 1876-1889

\bibitem{mac2006} J. M. McCollum, G. D. Peterson, C. D. Cox, M.
L. Simpson, N. F. Samatova. The sorting direct method for stochastic
simulation of biochemical systems with varying reaction execution
behavior. Comput. Bio. Chem. 30 (2006) 39\textendash{}49

\bibitem{petzold2006} H. Li, L. R. Petzold. Logarithmic Direct Method
for Discrete Stochastic Simulation of Chemically Reacting Systems.
Technical report (2006) 

\bibitem{gillespie2003} D. T. Gillespie, L. R. Petzold. Improved leap-size
selection for accelerated stochastic simulation. J. Chem. Phys. 119
(2003) 8229\textendash{}8234

\bibitem{chat2005} A. Chatterjee, D. Vlachos, M. Katsoulakis. Binomial
distribution based \textgreek{t}-leap accelerated stochastic simulation.
J. Chem. Phys. 122 (2005) 024112

\bibitem{cao2006} Y Cao, D. T. Gillespie, L. R. Petzold. Efficient
stepsize selection for the tau-leaping simulation method. J. Chem.
Phys. 124 (2006) 044109

\bibitem{tian2006} T. Tian, K. Burragex. Binomial leap methods
for simulating stochastic chemical kinetics. J. Chem. Phys. 121 (2006)
10356\textendash{}10364

\bibitem{cao2005} Y. Cao, L. R. Petzold. Trapezoidal tau-leaping
formula for the stochastic simulation of chemically reacting systems.
Proc. Found. Syst. Biol. Eng. (FOSBE 2005), pp. 149\textendash{}52

\bibitem{hasel2002} E. L. Haseltine, J. B. Rawlings. Approximate
simulation of coupled fast and slow reactions for stochastic chemical
kinetics. J. Chem. Phys. 117 ( 2002) 6959\textendash{}6969

\bibitem{rath2003} M. Rathinam, L. R. Petzold, Y. Cao, D. T. Gillespie.
Stiffness in stochastic chemically reacting systems: the implicit
tau-leaping method. J. Chem. Phys. 119 (2003) 12784\textendash{}94

\bibitem{samant2005} A. Samant, D. G. Vlachos. Overcoming stiffness
in stochastic simulation stemming from partial equilibrium: a multiscale
Monte Carlo algorithm. J. Chem. Phys. 123 (2005) 144114


\bibitem{cao2005b} Y. Cao, D. T. Gillespie, L. R. Petzold. Avoiding
negative populations in explicit Poisson tau-leaping. J. Chem. Phys.
123 (2005) 054104


\bibitem{tanaka2004} T. Tanaka, K. Oka, S. Ohnuki, S. Yamashita, T.
Suda, S. Watanabe, E. Wakai. Synergistic effect of helium and hydrogen
for defect evolution under multi-ion irradiation of Fe\textendash{}Cr
ferritic alloys. J. of Nucl. Mater. 329\textendash{}333 (2004) 294-298 

\bibitem{martinez2011} E. Martinez, P. R. Monasterio and J. Marian.
Billion-atom Synchronous Parallel Kinetic Monte Carlo Simulations
of Critical 3D Ising Systems, J Comput Phys. 230 (2011) 1359-1369





\end{thebibliography}
\end{document}